\documentclass[aps,twocolumn]{revtex4}%
\usepackage{amsfonts}
\usepackage{amsmath}
\usepackage{amssymb}
\usepackage{appendix}
\usepackage{graphicx}%
\providecommand{\U}[1]{\protect\rule{.1in}{.1in}}
\providecommand{\U}[1]{\protect\rule{.1in}{.1in}}

\begin{document}
\title{Exact Floquet solutions of quantum driven systems}
\author{\ Xiao-Bo Yan}
\email{xiaoboyan@126.com}
\affiliation{School of Physics and Electronic Engineering, Northeast Petroleum University, Daqing, 163318, China}
\date{\today}

\begin{abstract}
How to accurately solve time-dependent Schr\"odinger equation is an interesting and important problem. Here, we propose a novel method to obtain the \textit{exact Floquet solutions} of the Schr\"odinger equation for periodically driven systems by using  Floquet theorem and a time-dependent unitary transformation. Using the method, we give out the exact Floquet solutions of wave function for three interesting physical models---linear potential model, harmonic oscillator model, and the coupled harmonic oscillator model in the presence of a periodic driving. In addition to the quasienergy, we also give out the analytic expression of Berry phase for the harmonic oscillator model. Moreover, the idea presented in this paper can be used in mathematics to solve partial differential equations.
\end{abstract}
\maketitle

\section{Introduction}

In the past few decades, the study of the exact analytical solution for time-dependent schr\"odinger equation has been an interesting subject. Many efforts have been invested to the subject \cite{Ray1982pra,Dantas1992pra,Truscott1993prl,Yeon1994pra,Ji1995PRA4268,Ji1995PRA3352,Feng1995pla,Wang1995pra,Cocke1996pra,Guedes2001pra,Feng2001pra,Pedrosa1997pra}. For example, in Refs. \cite{Ji1995PRA3352,Pedrosa1997pra} the exact wave function of the harmonic oscillator with time-dependent frequency and mass has been obtained, and in Refs. \cite{Feng1995pla,Wang1995pra}, the authors studied the exact solution for the motion of a particle in a Paul trap \cite{Paul1990pmp}.
Besides the interest in mathematics, these problems connects with various applications to
many physical problems, such as the application in the study of Berry phase of a quantum state \cite{Ji1995PRA3352}, and the application to describe the motion of charged particles in an electromagnetic field \cite{Paul1990pmp,Cook1985pra}.

A periodically driven system is a kind of special and important time-dependent system, whose Hamiltonian is generally periodic in time. It is not possible to find energy eigenstates of the system because the energy is not conserved \cite{Yan2021ejp}. While we can obtain the quasienergy eigenstates of the system according to the Floquet theorem \cite{Shirley1965pr,Buchleitner2002PR,Eckardt2015NJP,Guo2020NJP}. Recently, these kinds of systems have attracted much attention \cite{ Grozdanov1988pra,Sambe1973PRA,Sacha2015pra,Else2016prl,Guo2020NJP,Casas2001jpa,Sacha2018RPP,Giergiel2019NJP,Sacha2020book,Matus2021pra,Junk2020prb}. For example, the periodically driven particle bouncing in the gravitational field, and the periodically driven harmonic oscillator have been investigated in the study of time crystals \cite{Sacha2015pra,Sacha2018RPP,Giergiel2019NJP,Sacha2020book}. Usually, in addition to the perturbation theory \cite{Sambe1973PRA}, the common method to study periodic systems is the action-angle variables method \cite{Buchleitner2002PR,Sacha2020book}. 
However, some approximations are often required in the study of periodically driven systems using the action-angle variables method, such as the secular approximation \cite{Buchleitner2002PR} and the high-frequency approximation \cite{Eckardt2015NJP}.

Here, we give a novel and concise method to exactly solve the time-dependent Schr\"odinger equation for periodically driven systems. For these systems, the problem of solving schr\"odinger equation can be  turned into that of solving the eigen equation of Floquet Hamiltonian according to the Floquet theorem. The central idea of the method is to separate space-time coordinates in the Floquet Hamiltonian by applying a time-dependent unitary transformation, resulting that the transformed Floquet Hamiltonian just consists of the Hamiltonian of a undriven system (or a solvable system) and a differential equation that only depends on time. According to the simple form of the transformed  Floquet Hamiltonian, the exact solution of the time-dependent Schr\"odinger equation can be easily obtained. As we will see, besides simple systems, this method can be applied to more complex systems, such as multi-driving systems and coupled systems. One advantage of this method is that the exact Floquet solution of wave function has a concise form. Another advantage is that from the exact  solution, we can clearly see the relationship between the wave function of the system and that of the undriven system.

\section{Models and equations}

The Floquet systems with periodic Hamiltonian $\hat{H}(t)=\hat{H}(t+T)$ enables us to use the Floquet theorem, which says that we can expand any solution of the time-dependent Schr\"odinger equation in the form
\begin{eqnarray}
\psi(x,t)=\sum_{n}c_{n}e^{-\frac{i}{\hbar}\mathcal{E}_{n}t}u_{n}(x,t), 
\end{eqnarray}
with the periodicity condition $u_{n}(x,t)=u_{n}(x,t+T)$.
Here, $\mathcal{E}_{n}$ is the quasienergy of the system, and $u_{n}(x,t)$ are the eigenstates of the so-called Floquet Hamiltonian $\hat{H}_{F}=\hat{H}(t)-i\hbar\partial_{t}$, i.e.,
\begin{eqnarray}
\hat{H}_{F}u_{n}(x,t)=\mathcal{E}_{n}u_{n}(x,t),
\end{eqnarray}
It means that if we can obtain the eigenvalue $\mathcal{E}_{n}$ and eigenstate $u_{n}(x,t)$ of the Floquet Hamiltonian $\hat{H}_{F}$ in Eq. (2), the wave function of the system can be obtained.  In the following, We will use the Floquet theorem to respectively solve the  Schr\"odinger equations for the linear potential model, the harmonic oscillator model, and the coupled harmonic oscillator model in the presence of a periodic driving.

\subsection{Linear potential model}

The model of a driven particle in linear potential is just the physical model describing a driven particle bouncing in the gravitational field where the potential energy of the particle with mass $m$ is $mgx$. If the particle is periodically driven via the interaction term $\lambda x\cos(\omega t)$, then the Hamiltonian of the system is
\begin{eqnarray}
\hat{H}=\frac{\hat{p}^{2}}{2m}+mgx+\lambda x\cos(\omega t),\quad x\geq0.  
\end{eqnarray}%
Here, $\lambda$ is the driving strength and $\omega$ is the frequency of the driving field, and we have taken the coordinate representation where the coordinate operator $\hat{x}$ is the coordinate $x$ itself \cite{Yan2021ejp}.   The Floquet Hamiltonian is
\begin{eqnarray}
\hat{H}_{F}=\frac{\hat{p}^{2}}{2m}+mgx+\lambda x\cos(\omega t)-i\hbar\frac{\partial}{\partial t},  
\end{eqnarray}%
and its eigen equation is
\begin{eqnarray}
\hat{H}_{F}u_{n}(x,t)=\mathcal{E}_{n}u_{n}(x,t). 
\end{eqnarray}%

To solve Eq. (5), we first perform a unitary transformation as 
\begin{eqnarray}
u_{n}(x,t)=\hat{U}\phi_{n}(x,t),
\end{eqnarray}
with
\begin{eqnarray}
\hat{U}=e^{i\alpha A(t)x}e^{i\beta B(t)\hat{p}}.  
\end{eqnarray}%
Here, $A(t)$ and $B(t)$ are real functions, $\alpha$ and $\beta$ are real constants.
Under the unitary transformation, the eigen equation in Eq. (5) becomes
\begin{eqnarray}
\hat{H}_{F}^{\prime}\phi_{n}(x,t)=\mathcal{E}_{n}\phi_{n}(x,t), 
\end{eqnarray}%
with
\begin{eqnarray}
\hat{H}_{F}^{\prime}=\hat{U}^{\dagger}\hat{H}_{F}\hat{U}.
\end{eqnarray}%
After calculation, the  transformed Floquet Hamiltonian $H_{F}^{\prime}$ can be obtained as
\begin{eqnarray}
\hat{H}_{F}^{\prime}&=&\frac{\hat{p}^{2}}{2m}+\frac{\alpha\hbar A(t)\hat{p}}{m}+\frac{\alpha^{2}\hbar^{2}A^{2}(t)}{2m}+mgx-i\hbar\frac{\partial}{\partial t}\notag\\
&+&\lambda x\cos(\omega t)-mg\beta\hbar B(t)-\lambda\beta\hbar B(t)\cos(\omega t)\notag\\
&+&\alpha\hbar x\dot{A}(t)-\alpha\beta\hbar^{2}B(t)\dot{A}(t)+\beta\hbar\dot{B}(t)\hat{p}.
\end{eqnarray}%

In order to make the transformed Hamiltonian $H_{F}^{\prime}$ as simple as possible, we can separate the space-time coordinates by setting
\begin{eqnarray}
\lambda\cos(\omega t)+\alpha\hbar \dot{A}(t)=0,\notag\\
\frac{\alpha\hbar A(t)}{m}+\beta\hbar\dot{B}(t)=0,
\end{eqnarray}%
which can be satisfied as
\begin{eqnarray}
A(t)&=&\sin(\omega t),\quad \alpha=-\frac{\lambda}{\hbar \omega},\notag\\
B(t)&=&\cos(\omega t),\quad \beta=-\frac{\lambda}{m\hbar\omega^{2}}.
\end{eqnarray}%
Under the conditions in Eq. (12), the transformed Floquet Hamiltonian $H_{F}^{\prime}$ in Eq. (10) becomes 
\begin{eqnarray}
\hat{H}_{F}^{\prime}&=&\hat{H}_{0}+\hat{H}_{t},\\
\hat{H}_{0}&=&\frac{\hat{p}^{2}}{2m}+mgx,\\
\hat{H}_{t}&=&\frac{\lambda^{2}\sin^{2}(\omega t)}{2m\omega^{2}}+\frac{g\lambda\cos(\omega t)}{\omega^{2}}-i\hbar\frac{\partial}{\partial t}.
\end{eqnarray}%
From Eq. (13), we know that the eigenvalue of the Floquet Hamiltonian $H_{F}^{\prime}$ is the sum of the eigenvalues of the operators $\hat{H}_{0}$ and $\hat{H}_{t}$ i.e., $\mathcal{E}_{n}=E_{n}+E_{t}$, and the eigenstate $\phi_{n}(x,t)$ is the product of the eigenstates of the operators $\hat{H}_{0}$ and $\hat{H}_{t}$, i.e., $\phi_{n}(x,t)=\phi_{n}(x)\phi(t)$. The eigenstate of $\hat{H}_{0}$ is the well-known Airy function, i.e.,
\begin{eqnarray} \phi_{n}(x)=\mathcal{A}_{n}\cdot\mathrm{Ai}[(\frac{2}{mg^{2}\hbar^{2}})^{\frac{1}{3}}(mgx-E_{n})], 
\end{eqnarray}
here, $\mathcal{A}_{n}$ is the normalization constant which can be obtained by numerical methods, and the eigenvalue $E_{n}$ can be obtained as
\begin{eqnarray}
E_{n}=-(\frac{mg^{2}\hbar^{2}}{2})^{\frac{1}{3}}a_{n},
\end{eqnarray}
where $a_{n}$ is the nth zero of the Airy function.

Now let's solve for the eigenvalue $E_{t}$ and eigenstate $\phi(t)$ of the operator $\hat{H}_{t}$. From Eq. (15), we have
\begin{eqnarray}
\{\frac{\lambda^{2}\sin^{2}(\omega t)}{2m\omega^{2}}+\frac{g\lambda\cos(\omega t)}{\omega^{2}}-i\hbar\frac{\partial}{\partial t}\}\phi(t)=E_{t}\phi(t),
\end{eqnarray}%
from which the eigenvalue $E_{t}$ and eigenstate $\phi(t)$ can be obtained  respectively as
\begin{eqnarray}
E_{t}&=&\frac{\lambda^{2}}{4m\omega^{2}},\\
\phi(t)&=&e^{\frac{i\lambda(\lambda\cos(\omega t)-4mg)\sin(\omega t)}{4m\omega^{3}\hbar}},
\end{eqnarray}%
with the periodicity condition of $\phi(t)=\phi(t+T)$.
In fact, the eigenvalue $E^{\prime}_{t}=\frac{\lambda^{2}}{4m\omega^{2}}+s\hbar\omega$ and the eigenstate $\phi^{\prime}(t)=e^{i[s\omega t+\frac{\lambda(\lambda\cos(\omega t)-4mg)\sin(\omega t)}{4m\omega^{3}\hbar}]}$ with any integer $s$ is also a solution of equation (18). While they are equivalent because we have $e^{-\frac{i}{\hbar}E_{t}t}\phi(t)=e^{-\frac{i}{\hbar}E^{\prime}_{t}t}\phi^{\prime}(t)$ in the wave function $\psi(x,t)$ in Eq. (1). Hence, we can take $s=0$ for simplicity.

According to Eq. (1) and the results above, the wave function of the system can be given as
\begin{eqnarray}
\psi(x,t)=\sum_{n}c_{n}\mathcal{A}_{n}e^{-\frac{i}{\hbar}\mathcal{E}_{n}t}e^{-\frac{i\lambda\sin(\omega t)x}{\hbar\omega}}e^{-\frac{i\lambda\cos(\omega t)\hat{p}}{m\hbar\omega^{2}}}\times\notag\\ 
\mathrm{Ai}[(\frac{2}{mg^{2}\hbar^{2}})^{\frac{1}{3}}(mgx-E_{n})]e^{\frac{i\lambda(\lambda\cos(\omega t)-4mg)\sin(\omega t)}{4m\omega^{3}\hbar}},\quad 
\end{eqnarray}
here,
\begin{eqnarray}
\mathcal{E}_{n}=-(\frac{mg^{2}\hbar^{2}}{2})^{\frac{1}{3}}a_{n}+\frac{\lambda^{2}}{4m\omega^{2}}.
\end{eqnarray}

In fact, in addition to the single-driving system as above, the method can also be used to solve the Schr\"odinger equation for systems with two or more driving fields. Now assume that the model above is driven by two driving fields, and then the Hamiltonian in Eq. (3) becomes
\begin{eqnarray}
\hat{H}=\frac{\hat{p}^{2}}{2m}+mgx+x(\lambda_{1}\cos(\omega_{1}t)+\lambda_{2}\cos(\omega_{2}t)).  
\end{eqnarray}
After a similar calculation, we can obtain the appropriate unitary transformation as 
\begin{eqnarray}
\hat{U}=e^{i(\alpha_{1}A_{1}(t)+\alpha_{2}A_{2}(t))x}e^{i(\beta_{1}B_{1}(t)+\beta_{2}B_{2}(t))\hat{p}}
\end{eqnarray}
with
\begin{eqnarray}
A_{1}(t)&=&\sin(\omega_{1}t),\quad A_{2}(t)=\sin(\omega_{2}t),\notag\\
B_{1}(t)&=&\cos(\omega_{1}t),\quad
B_{2}(t)=\cos(\omega_{2}t),\notag\\
\alpha_{1}&=&-\frac{\lambda_{1}}{\hbar\omega_{1}},\quad\quad 
\beta_{1}=-\frac{\lambda_{1}}{m\hbar\omega_{1}^{2}},\notag\\
\alpha_{2}&=&-\frac{\lambda_{2}}{\hbar\omega_{2}},\quad\quad  \beta_{2}=-\frac{\lambda_{2}}{m\hbar\omega_{2}^{2}}.
\end{eqnarray}%
With the unitary transformation in Eq. (24), we can obtain the wave function of the system as
\begin{eqnarray}
\psi(x,t)=\sum_{n}c_{n}\mathcal{A}_{n}e^{-\frac{i}{\hbar}\mathcal{E}_{n}t}e^{-i\{\frac{\lambda_{1}\sin(\omega_{1}t)}{\hbar\omega_{1}}+\frac{\lambda_{2}\sin(\omega_{2}t)}{\hbar\omega_{2}}\}x}\times\notag\\
e^{-i\{\frac{\lambda_{1}\cos(\omega_{1}t)}{m\hbar\omega_{1}^{2}}+\frac{\lambda_{2}\cos(\omega_{2}t)}{m\hbar\omega_{2}^{2}}\}\hat{p}}\mathrm{Ai}[(\frac{2}{mg^{2}\hbar^{2}})^{\frac{1}{3}}(mgx-E_{n})]\times\notag\\
e^{\frac{i\lambda_{2}}{8\hbar}\{\frac{4\lambda_{1}}{m\omega_{1}\omega_{2}}(\frac{\sin(\omega_{1}+\omega_{2})t}{\omega_{1}+\omega_{2}}-\frac{\sin(\omega_{1}-\omega_{2})t}{\omega_{1}-\omega_{2}})+\frac{\lambda_{2}\sin(2\omega_{2}t)}{m\omega_{2}^{3}}-\frac{8g\sin(\omega_{2}t)}{\omega_{2}^{3}}\}}\notag\\ 
\times e^{\frac{i\lambda_{1}(\lambda_{1}\cos(\omega_{1} t)-4mg)\sin(\omega_{1}t)}{4m\omega_{1}^{3}\hbar}},\qquad
\end{eqnarray}
here,
\begin{eqnarray}
\mathcal{E}_{n}=-(\frac{mg^{2}\hbar^{2}}{2})^{\frac{1}{3}}a_{n}+\frac{\lambda_{1}^{2}}{4m\omega_{1}^{2}}+\frac{\lambda_{2}^{2}}{4m\omega_{2}^{2}}.
\end{eqnarray}
It is easy to verify that the wave function in Eq. (26) will be consistent with the result in Eq. (21) if we take $\lambda_{2}=0$.

\subsection{Harmonic oscillator model}

Now we study the harmonic oscillator with frequency $\omega_{m}$ and mass $m$, which is driven by a periodic driving term $\lambda x\cos(\omega t)$. The Hamiltonian of this model is 
\begin{eqnarray}
\hat{H}=\frac{\hat{p}^{2}}{2m}+\frac{1}{2}m\omega_{m}^{2}x^{2}+\lambda x\cos(\omega t).
\end{eqnarray}%
The corresponding Floquet Hamiltonian is
\begin{eqnarray}
\hat{H}_{F}=\frac{\hat{p}^{2}}{2m}+\frac{1}{2}m\omega_{m}^{2}x^{2}+\lambda x\cos(\omega t)-i\hbar\frac{\partial}{\partial t},  
\end{eqnarray}%
and its eigen equation is
\begin{eqnarray}
\hat{H}_{F}u_{n}(x,t)=\mathcal{E}_{n}u_{n}(x,t). 
\end{eqnarray}%
We take the same form of unitary transformation as that in Eq. (7), and after calculation we can obtain the transformed Hamiltonian $H_{F}^{\prime}$ as
\begin{eqnarray}
\hat{H}_{F}^{\prime}&=&\hat{U}^{\dagger}\hat{H}_{F}\hat{U}\notag\\
&=&\frac{\hat{p}^{2}}{2m}+\frac{\alpha\hbar A(t)\hat{p}}{m}+\frac{\alpha^{2}\hbar^{2}A^{2}(t)}{2m}+\frac{1}{2}m\omega_{m}^{2}x^{2}\notag\\
&+&\lambda x\cos(\omega t)-m\omega_{m}^{2}x\beta\hbar B(t)+\alpha\hbar x\dot{A}(t)\notag\\
&+&\frac{1}{2}m\omega_{m}^{2}\beta^{2}\hbar^{2}B^{2}(t)-\alpha\beta\hbar^{2}\dot{A}(t)B(t)+\beta\hbar\dot{B}(t)\hat{p}\notag\\
&-&\lambda\beta\hbar B(t)\cos(\omega t)-i\hbar\frac{\partial}{\partial t}.
\end{eqnarray}%
If we take
\begin{eqnarray}
A(t)&=&\sin(\omega t),\quad \alpha=\frac{\lambda\omega}{\hbar(\omega_{m}^{2}-\omega^{2})},\notag\\
B(t)&=&\cos(\omega t),\quad \beta=\frac{\lambda}{m\hbar(\omega_{m}^{2}-\omega^{2})},
\end{eqnarray}%
the transformed Hamiltonian $\hat{H}_{F}^{\prime}$ in Eq. (31) will become very simple, i.e.,
\begin{eqnarray}
\hat{H}_{F}^{\prime}&=&\hat{H}_{0}+\hat{H}_{t},\\
\hat{H}_{0}&=&\frac{\hat{p}^{2}}{2m}+\frac{1}{2}m\omega_{m}^{2}x^{2},\notag\\
\hat{H}_{t}&=&\frac{\lambda^{2}(\omega^{2}\sin^{2}(\omega t)-\omega_{m}^{2}\cos^{2}(\omega t))}{2m(\omega_{m}^{2}-\omega^{2})^{2}}-i\hbar\frac{\partial}{\partial t}\notag.
\end{eqnarray}%

The Hamiltonian $\hat{H}_{0}$ in Eq. (33) is the free Hamiltonian of a quantum harmonic oscillator, and its eigenvalue $E_{n}$ and eigenstate $\phi_{n}(x)$ respectively are the well-known $(n+\frac{1}{2})\hbar\omega_{m}$ and $N_{n}e^{-\frac{a^{2}x^{2}}{2}}H_{n}(ax)$, here the normalization constant $N_{n}=(\frac{a}{\sqrt{\pi}2^{n}n!})^{\frac{1}{2}}$, $a=\sqrt{\frac{m\omega_{m}}{\hbar}}$ and $H_{n}(ax)$ is the Hermite polynomial.
The eigenvalue $E_{t}$ and eigenstate $\phi(t)$ of operator $\hat{H}_{t}$ can be easily obtained as
\begin{eqnarray}
E_{t}&=&\frac{\lambda^{2}}{4m(\omega^{2}-\omega_{m}^{2})},\\
\phi(t)&=&e^{\frac{i\lambda^{2}(\omega^{2}+\omega_{m}^{2})\sin(2\omega t)}{8m\hbar\omega(\omega^{2}-\omega_{m}^{2})^{2}}},
\end{eqnarray}
with the periodicity condition of $\phi(t)=\phi(t+T)$.
Finally, we can obtain the wave function of the system as 
\begin{eqnarray}
\psi(x,t)=\sum_{n}c_{n}N_{n}e^{-\frac{i}{\hbar}\mathcal{E}_{n}t}e^{i\frac{\lambda\omega\sin(\omega t)}{\hbar(\omega_{m}^{2}-\omega^{2})}x}e^{i\frac{\lambda\cos(\omega t)}{m\hbar(\omega_{m}^{2}-\omega^{2})}\hat{p}}\notag\\
\times e^{-\frac{a^{2}x^{2}}{2}}H_{n}(ax)e^{\frac{i\lambda^{2}(\omega^{2}+\omega_{m}^{2})\sin(2\omega t)}{8m\hbar\omega(\omega^{2}-\omega_{m}^{2})^{2}}},\quad 
\end{eqnarray}
here,
\begin{eqnarray}
\mathcal{E}_{n}=(n+\frac{1}{2})\hbar\omega_{m}+\frac{\lambda^{2}}{4m(\omega^{2}-\omega_{m}^{2})}
\end{eqnarray}
according to $\mathcal{E}_{n}=E_{n}+E_{t}$.

From the wave function $\psi(x,t)$ in Eq. (36), we can obtain the analytic expression of Berry phase in the system. 
Choosing an initial state makes the evolution cyclic, which means
\begin{eqnarray}
\psi(x,T)=e^{i\chi}\psi(x,0),
\end{eqnarray}
where $\chi$ is the overall phase. For simplicity, we assume that the system is in the state $\psi(x,t)$ with $c_{n}=1$, and
from Eqs. (36)--(38), the overall phase can be given as
\begin{eqnarray}
\chi&=&-\frac{\mathcal{E}_{n}\cdot T}{\hbar}\notag\\
&=&-((n+\frac{1}{2})\omega_{m}+\frac{\lambda^{2}}{4m\hbar(\omega^{2}-\omega_{m}^{2})})T.
\end{eqnarray}
As is well-known, the overall phase can be split into two parts, namely the dynamical phase $\delta$ and Berry phase $\gamma$.
The dynamic phase  $\delta$ can be defined as
\begin{eqnarray}
\delta=-\frac{1}{\hbar}\int_{0}^{T}\int_{-\infty}^{\infty}\psi^{*}(x,t)\hat{H}(t)\psi(x,t)dxdt.
\end{eqnarray}
Then, the Berry phase  $\gamma$ is equal to the subtraction of the overall phase and the dynamical phase, i.e.,
\begin{eqnarray}
\gamma=\chi-\delta,
\end{eqnarray} 
which can be obtained as
\begin{eqnarray}
\gamma=\frac{\omega\lambda^{2}\pi}{m\hbar(\omega^{2}-\omega_{m}^{2})^{2}}
\end{eqnarray} 
according to Eqs. (36)--(42).

\subsection{Coupled Harmonic oscillator model}

In general, the time-dependent Schr\"odinger equation for coupled systems is very difficult to solve. In this section, we exactly solve the Schr\"odinger equation for coupled harmonic oscillator system with the method mentioned above. The system consists of two harmonic oscillators with masses $m_{1}$ and $m_{2}$, and frequencies $\omega_{1}$ and $\omega_{2}$ respectively. The two harmonic oscillators are coupled to each other via the interaction term $gx_{1}x_{2}$ with the interaction strength $g$, and the first harmonic oscillator is driven by the driving term $\lambda x_{1}\cos(\omega t)$. Then the whole Hamiltonian of the system is
\begin{eqnarray}
\hat{H}&=&\frac{\hat{p}^{2}_{1}}{2m_{1}}+\frac{1}{2}m_{1}\omega_{1}^{2}x_{1}^{2}+\frac{\hat{p}^{2}_{2}}{2m_{2}}+\frac{1}{2}m_{2}\omega_{2}^{2}x_{2}^{2}\notag\\
&+&gx_{1}x_{2}+\lambda x_{1}\cos(\omega t).
\end{eqnarray}
The corresponding Floquet Hamiltonian is 
\begin{eqnarray}
\hat{H}_{F}&=&\frac{\hat{p}^{2}_{1}}{2m_{1}}+\frac{1}{2}m_{1}\omega_{1}^{2}x_{1}^{2}+\frac{\hat{p}^{2}_{2}}{2m_{2}}+\frac{1}{2}m_{2}\omega_{2}^{2}x_{2}^{2}\notag\\
&+&gx_{1}x_{2}+\lambda x_{1}\cos(\omega t)-i\hbar\frac{\partial}{\partial t},
\end{eqnarray}
and its eigen equation is
\begin{eqnarray}
\hat{H}_{F}u_{n}(x,t)=\mathcal{E}_{n}u_{n}(x,t). 
\end{eqnarray}%
To solve Eq. (45), we take a unitary transformation as
\begin{eqnarray}
\hat{U}=e^{i\alpha_{1}A_{1}(t)x_{1}}e^{i\beta_{1}B_{1}(t)\hat{p}_{1}}e^{i\alpha_{2}A_{2}(t)x_{2}}e^{i\beta_{2}B_{2}(t)\hat{p}_{2}}, 
\end{eqnarray}%
under which the Floquet Hamiltonian becomes
\begin{eqnarray}
\hat{H}_{F}^{\prime}&=&\hat{U}^{\dagger}\hat{H}_{F}\hat{U}\notag\\
&=&\frac{\hat{p}^{2}_{1}}{2m_{1}}+\frac{\hbar\alpha_{1}A_{1}(t)\hat{p}_{1}}{m_{1}}+\frac{\alpha_{1}^{2}\hbar^{2}A_{1}^{2}(t)}{2m_{1}}+\frac{1}{2}m_{1}\omega_{1}^{2}x_{1}^{2}\notag\\
&-&m_{1}\omega_{1}^{2}\hbar\beta_{1}B_{1}(t)x_{1}+\frac{1}{2}m_{1}\omega_{1}^{2}\beta_{1}^{2}\hbar^{2}B_{1}^{2}(t)+gx_{1}x_{2}\notag\\
&-&gx_{1}\hbar\beta_{2}B_{2}(t)-gx_{2}\hbar\beta_{1}B_{1}(t)+\lambda x_{1}\cos(\omega t)\notag\\
&+&g\beta_{1}\beta_{2}\hbar^{2}B_{1}(t)B_{2}(t)-\lambda\hbar\beta_{1}B_{1}(t)\cos(\omega t)+\frac{\hat{p}^{2}_{2}}{2m_{2}}\notag\\
&+&\frac{\hbar\alpha_{2}A_{2}(t)\hat{p}_{2}}{m_{2}}+\frac{\alpha_{2}^{2}\hbar^{2}A_{2}^{2}(t)}{2m_{2}}-m_{2}\omega_{2}^{2}x_{2}\hbar\beta_{2}B_{2}(t)\notag\\
&+&\frac{1}{2}m_{2}\omega_{2}^{2}x_{2}^{2}+\frac{1}{2}m_{2}\omega_{2}^{2}\beta_{2}^{2}\hbar^{2}B_{2}^{2}(t)+\hbar\beta_{1}\dot{B}_{1}(t)\hat{p}_{1}\notag\\
&+&\hbar\alpha_{1}\dot{A}_{1}(t)x_{1}-\hbar^{2}\alpha_{1}\beta_{1}\dot{A}_{1}(t)B_{1}(t)+\hbar\beta_{2}\dot{B}_{2}(t)\hat{p}_{2}\notag\\
&+&\hbar\alpha_{2}\dot{A}_{2}(t)x_{2}-\hbar^{2}\alpha_{2}\beta_{2}\dot{A}_{2}(t)B_{2}(t)-i\hbar\frac{\partial}{\partial t}.
\end{eqnarray}
If we take 
\begin{eqnarray}
A_{1}(t)=A_{2}(t)=\sin(\omega t),\notag\\  B_{1}(t)=B_{2}(t)=\cos(\omega t),
\end{eqnarray}
and
\begin{eqnarray}
\alpha_{1}=\frac{m_{1}m_{2}\lambda\omega(\omega^{2}-\omega_{2}^{2})}{[g^{2}-m_{1}m_{2}(\omega^{2}-\omega_{1}^{2})(\omega^{2}-\omega_{2}^{2})]\hbar},\notag\\
\alpha_{2}=\frac{gm_{2}\lambda\omega}{[g^{2}-m_{1}m_{2}(\omega^{2}-\omega_{1}^{2})(\omega^{2}-\omega_{2}^{2})]\hbar},\notag\\
\beta_{1}=\frac{m_{2}\lambda(\omega^{2}-\omega_{2}^{2})}{[g^{2}-m_{1}m_{2}(\omega^{2}-\omega_{1}^{2})(\omega^{2}-\omega_{2}^{2})]\hbar},\notag\\
\beta_{2}=\frac{g\lambda}{[g^{2}-m_{1}m_{2}(\omega^{2}-\omega_{1}^{2})(\omega^{2}-\omega_{2}^{2})]\hbar},
\end{eqnarray}
then, the transformed Floquet Hamiltonian in Eq. (47) becomes
\begin{eqnarray}
\hat{H}_{F}^{\prime}&=&\hat{H}_{0}+\hat{H}_{t},
\end{eqnarray}
here,
\begin{widetext}
    \begin{eqnarray}
	\hat{H}_{0}&=&\frac{\hat{p}^{2}_{1}}{2m_{1}}+\frac{1}{2}m_{1}\omega_{1}^{2}x_{1}^{2}+\frac{\hat{p}^{2}_{2}}{2m_{2}}+\frac{1}{2}m_{2}\omega_{2}^{2}x_{2}^{2}
	+gx_{1}x_{2},\notag\\
	\hat{H}_{t}&=&\frac{m_{2}\lambda^{2}[\omega^{2}(g^{2}+m_{1}m_{2}(\omega^{2}-\omega_{2}^{2})^{2})\sin^{2}(\omega t)-(m_{1}m_{2}\omega_{1}^{2}(\omega^{2}-\omega_{2}^{2})^{2}+g^{2}(2\omega^{2}-\omega_{2}^{2}))\cos^{2}(\omega t)]}{2[g^{2}-m_{1}m_{2}(\omega^{2}-\omega_{1}^{2})(\omega^{2}-\omega_{2}^{2})]^{2}}	-i\hbar\frac{\partial}{\partial t}.
	\end{eqnarray}
\end{widetext}
The eigenvalue $E_{n_{1},n_{2}}$ and eigenstate $\phi_{n_{1},n_{2}}(x_{1},x_{2})$ of $\hat{H}_{0}$ can be obtained, see the detailed calculations
in Appendix A. The eigenvalue $E_{n_{1},n_{2}}$ is given by
\begin{eqnarray}
E_{n_{1},n_{2}}=\hbar\Omega_{1}(n_{1}+\frac{1}{2})+\hbar\Omega_{2}(n_{2}+\frac{1}{2}),\notag\\
n_{1}, n_{2}=0, 1, 2, \dots
\end{eqnarray}
here, $\Omega_{1}$ and $\Omega_{2}$ are determined by 
\begin{eqnarray}
\Omega^{2}_{1}&=&(\omega_{1}^{2}\cos^{2}\theta+\omega_{2}^{2}\sin^{2}\theta)-\frac{g\sin 2\theta}{\sqrt{m_{1}m_{2}}},\\
\Omega^{2}_{2}&=&(\omega_{2}^{2}\cos^{2}\theta+\omega_{1}^{2}\sin^{2}\theta)+\frac{g\sin 2\theta}{\sqrt{m_{1}m_{2}}},
\end{eqnarray}
with
\begin{eqnarray}
\theta=\frac{1}{2}\arctan\frac{2g}{\sqrt{m_{1}m_{2}}(\omega_{2}^{2}-\omega_{1}^{2})}.
\end{eqnarray}
And the eigenstate $\phi_{n_{1},n_{2}}(x_{1},x_{2})$ of $\hat{H}_{0}$ is given by
\begin{eqnarray}
\phi_{n_{1},n_{2}}(x_{1},x_{2})=\mathcal{N}e^{-\frac{a_{1}^{2}X_{1}^{2}+a_{2}^{2}X_{2}^{2}}{2}}H_{n_{1}}(a_{1}X_{1})H_{n_{2}}(a_{2}X_{2}),\notag\\
\end{eqnarray}
here $H_{n_{i}}(a_{i}X_{i})$ is the Hermite polynomial, and
\begin{eqnarray}
X_{1}&=&\eta x_{1}\cos\theta-\eta^{-1}x_{2}\sin\theta,\\
X_{2}&=&\eta^{-1}x_{2}\cos\theta+\eta x_{1}\sin\theta,\\
\mathcal{N}&=&\sqrt{\frac{a_{1}a_{2}}{\pi 2^{n_{1}+n_{2}}n_{1}!n_{2}!}},\\
a_{i}&=&\sqrt{\frac{M\Omega_{i}}{\hbar}}, \end{eqnarray}
with $M=\sqrt{m_{1}m_{2}}$ and $\eta=\sqrt[4]{m_{1}/m_{2}}$.

According to Eq. (51) and the periodicity condition $\phi(t)=\phi(t+T)$,  the eigenvalue $E_{t}$ and eigenstate $\phi(t)$ of $H_{t}$ can be obtained respectively as
\begin{eqnarray}
E_{t}=-\frac{m_{2}\lambda^{2}(\omega^{2}-\omega_{2}^{2})}{4[g^{2}-m_{1}m_{2}(\omega^{2}-\omega_{1}^{2})(\omega^{2}-\omega_{2}^{2})]},
\end{eqnarray}
and 
\begin{eqnarray}
\phi(t)=e^{\frac{im_{2}\lambda^{2}[m_{1}m_{2}(\omega^{2}+\omega_{1}^{2})(\omega^{2}-\omega_{2}^{2})^{2}+g^{2}(3\omega^{2}-\omega_{2}^{2})]\sin(2\omega t)}{8\hbar\omega[g^{2}-m_{1}m_{2}(\omega^{2}-\omega_{1}^{2})(\omega^{2}-\omega_{2}^{2})]^{2}}}.\; 
\end{eqnarray}
It's easy to verify that the results in Eqs. (61) and (62) will be consistent with those in Eqs. (34) and (35) if we take $g=0$.

Hence, from Eqs. (52) and (61), the eigenvalue  $\mathcal{E}_{n_{1},n_{2}}$ of the Floquet Hamiltonian $\hat{H}_{F}$ in Eq. (44) can be obtained as
\begin{eqnarray}
\mathcal{E}_{n_{1},n_{2}}&=&\hbar\Omega_{1}(n_{1}+\frac{1}{2})+\hbar\Omega_{2}(n_{2}+\frac{1}{2})\notag\\
&-&\frac{m_{2}\lambda^{2}(\omega^{2}-\omega_{2}^{2})}{4[g^{2}-m_{1}m_{2}(\omega^{2}-\omega_{1}^{2})(\omega^{2}-\omega_{2}^{2})]}.
\end{eqnarray}
According to the above results, the wave function of the system can be obtained as
\begin{eqnarray}
\psi(x,t)=\sum_{n_{1},n_{2}}c_{n_{1},n_{2}}e^{-\frac{i}{\hbar}\mathcal{E}_{n_{1},n_{2}}t}\hat{U}\phi_{n_{1},n_{2}}(x_{1},x_{2})\phi(t),\quad 
\end{eqnarray}
which is too long to be explicitly written here.
After a similar calculation as that in Eqs. (38)--(42), and from Eq. (64), we can obtain the Berry phase of the system as
\begin{eqnarray}
\gamma=\frac{m_{2}\pi\lambda^{2}\omega(g^{2}+m_{1}m_{2}(\omega^{2}-\omega_{2}^{2})^{2})}{(g^{2}-m_{1}m_{2}(\omega^{2}-\omega_{1}^{2})(\omega^{2}-\omega_{2}^{2}))^{2}\hbar},
\end{eqnarray} 
which will be equal to that in Eq. (42) as $g=0$.

\section{Conclusion}

In summary, we propose a simple and interesting method to accurately solve time-dependent schr\"odinger equation by using the Floquet theorem and a time-dependent unitary transformation. Using the method, we can obtain the  exact Floquet solutions  of wave function and the quasienergies of three interesting physical systems. From the simple form of the Floquet solution of wave function, we can clearly see the relationship between wave function of the system and that of the undriven system. It can also be seen that the method is still suitable to solve the time-dependent schr\"odinger equations for coupled systems, as well as the systems with multiple driving fields. Actually, in addition to the several models in this paper, the method can also be used to solve other quantum systems, such as the driven quantum wells \cite{Wagner1996PRL}. We believe that the results in this paper are beneficial to many aspects of physics.

\begin{acknowledgements}
	We thank Dr. X. D. Tian for helpful discussions.
\end{acknowledgements}

\appendix{}
\section{The eigenvalue and eigenstate of $\hat{H}_{0}$}

In the absence of a driving field, the Hamiltonian of the coupled harmonic oscillator system is
\begin{eqnarray}
\hat{H}_{0}=\frac{\hat{p}^{2}_{1}}{2m_{1}}+\frac{1}{2}m_{1}\omega_{1}^{2}x_{1}^{2}+\frac{\hat{p}^{2}_{2}}{2m_{2}}+\frac{1}{2}m_{2}\omega_{2}^{2}x_{2}^{2}+gx_{1}x_{2}.\quad 
\end{eqnarray} 
To obtain the eigenvalue and eigenstate of Hamiltonian $\hat{H}_{0}$, we first do the substitution
\begin{eqnarray}
x_{1}&=&\eta^{-1}(X_{1}\cos\theta+X_{2}\sin\theta),\notag\\
x_{2}&=&\eta(X_{2}\cos\theta-X_{1}\sin\theta),\notag\\
\hat{p}_{1}&=&\eta(\hat{P}_{1}\cos\theta+\hat{P}_{2}\sin\theta),\notag\\
\hat{p}_{2}&=&\eta^{-1}(\hat{P}_{2}\cos\theta-\hat{P}_{1}\sin\theta).
\end{eqnarray}
According to the commutation relation between $x_{i}$ and $\hat{p}_{j}$, it can be easily proofed that $[X_{i}, P_{j}]=i\hbar\delta_{ij}$, $[X_{i}, X_{j}]=0$ and $[P_{i}, P_{j}]=0$ with $i, j=1, 2$.

Substituting Eq. (A2) into Eq. (A1), and setting
\begin{eqnarray}
\eta&=&\sqrt[4]{m_{1}/m_{2}},\\
\theta&=&\frac{1}{2}\arctan\frac{2g}{\sqrt{m_{1}m_{2}}(\omega_{2}^{2}-\omega_{1}^{2})},
\end{eqnarray}
the Hamiltonian $\hat{H}_{0}$ in Eq. (A1) becomes
\begin{eqnarray}
\hat{H}_{0}&=&\frac{\hat{P}_{1}^{2}}{2M}+\frac{1}{2}M\Omega^{2}_{1}X_{1}^{2}+\frac{\hat{P}_{2}^{2}}{2M}+\frac{1}{2}M\Omega^{2}_{2}X_{2}^{2},
\end{eqnarray}
with
\begin{eqnarray}
M&=&\sqrt{m_{1}m_{2}},\\
\Omega^{2}_{1}&=&(\omega_{1}^{2}\cos^{2}\theta+\omega_{2}^{2}\sin^{2}\theta)-\frac{g\sin 2\theta}{\sqrt{m_{1}m_{2}}},\\
\Omega^{2}_{2}&=&(\omega_{2}^{2}\cos^{2}\theta+\omega_{1}^{2}\sin^{2}\theta)+\frac{g\sin 2\theta}{\sqrt{m_{1}m_{2}}}.
\end{eqnarray}
It means that  the coupled harmonic oscillator system with different masses and frequencies can be regarded as consisting of two free harmonic oscillators with the same mass and different frequencies. Hence, the eigenvalue $E_{n_{1},n_{2}}$ of $\hat{H}_{0}$ is just the sum of eigenvalues of the Hamiltonian of the two free harmonic oscillators in Eq. (A5), see Eq. (52) in text, and the eigenstate $\phi_{n_{1},n_{2}}(x_{1},x_{2})$ of  $\hat{H}_{0}$ is just the product of eigenstates of Hamiltonian of the two free harmonic oscillators, see Eq. (56) in text.

\end{document}